\documentclass[arps,twocolumn,showpacs]{revtex4}

\input{epsf}
\newcommand\lsim{\mathrel{\rlap{\lower4pt\hbox{\hskip1pt$\sim$}}
        \raise1pt\hbox{$<$}}}
\newcommand\gsim{\mathrel{\rlap{\lower4pt\hbox{\hskip1pt$\sim$}}
        \raise1pt\hbox{$>$}}}
\def\thetaB{\mbox{\boldmath$\hat\theta$}}

\begin{document}

\title{Magnification--Temperature Correlation: the Dark Side of ISW Measurements}
\author{Marilena LoVerde$^{1,2}$, Lam Hui$^{1,2}$, Enrique Gazta\~{n}aga$^{3,4}$}

\affiliation{
$^{1}$Institute for Strings, Cosmology and Astro-particle Physics (ISCAP)\\
$^{2}$Department of Physics, Columbia University, New York, NY 10027\\
$^{3}$Institut de Ci\`encies de l'Espai, CSIC/IEEC, Campus UAB,
F. de Ci\`encies, Torre C5 par-2,  Barcelona 08193, Spain\\
$^{4}$INAOE, Astrof\'{\i}sica, Tonantzintla, Puebla 7200, Mexico\\   
marilena@phys.columbia.edu, lhui@astro.columbia.edu, gazta@aliga.ieec.uab.es
}
\date{\today}
\begin{abstract}
Integrated Sachs-Wolfe (ISW) measurements, which involve cross-correlating the 
microwave background anisotropies with the foreground large-scale structure 
(e.g. traced by galaxies/quasars), have proven to be an interesting probe of 
dark energy. We show that magnification bias, which is the inevitable modulation
of the foreground number counts by gravitational lensing, alters both the scale 
dependence and amplitude of the observed ISW signal. This is true especially at 
high redshifts because (1) the intrinsic galaxy-temperature signal diminishes
greatly back in the matter dominated era, (2) the lensing efficiency increases 
with redshift and (3) the number count slope generally steepens with redshift
in a magnitude limited sample. At $z \gsim 2$, the magnification-temperature 
correlation dominates over the intrinsic galaxy-temperature correlation and 
causes the observed ISW signal to increase with redshift, despite dark energy 
subdominance -- a result of the fact that magnification probes structures all 
the way from the observer to the sources. Ignoring magnification bias therefore 
can lead to (significantly) erroneous conclusions about dark energy. While the lensing modulation 
opens up an interesting high $z$ window for ISW measurements, high redshift
measurements are not expected to add much new information to low redshift ones 
if dark energy is indeed the cosmological constant. This is because lensing 
introduces significant covariance across redshifts. The most compelling reasons 
for pursuing high redshift ISW measurements are to look for potential surprises 
such as early dark energy domination or signatures of modified gravity.
We conclude with a discussion of existing measurements, the highest redshift of 
which is at the margin of being sensitive to the magnification effect. We also 
develop a formalism which might be of more general interest: to predict biases 
in estimating parameters when certain physical effects are ignored in interpreting 
observations.
\end{abstract}
\maketitle

\section{Introduction}
\label{intro}

In a universe with accelerating expansion, the gravitational potentials associated with large--scale structures decay. A photon traveling through a decaying potential will experience a net change in energy. This leads to a secondary anisotropy in the cosmic microwave background (CMB) called the integrated Sachs--Wolfe (ISW) effect \cite{SW67,KS85}. The ISW effect has been proposed as a tool to examine dark energy \cite{CT96}. Recent studies have detected the ISW signature providing an additional confirmation of the presence of dark energy \cite{ALS04,BC04,FGC03,FG04,N03,Scranton03,CGMFS06,QSOISW,PBM06}.  

Since inhomogeneities in the matter distribution trace gravitational potentials, one expects a significant correlation 
between the CMB temperature anisotropies and the distribution of large--scale structure at redshifts
where there is a non--negligible fraction of dark energy.  A universe with non--negligible curvature or contributions from anything other than pressureless matter can do the same. Bearing in mind the evidence for spatial flatness \cite{maxlrg,wmap}, we will only consider dark energy. 

The ISW anisotropy can be isolated from the primordial anisotropies in the CMB by cross--correlating the temperature and galaxy/quasar fluctuations. (Hereafter, 
galaxy and quasar can be considered roughly synonymous: essentially all statements about one apply to the other.)
Cross-correlating the temperature anisotropy in direction $\thetaB$, $\delta_T (\thetaB) =
\delta T(\thetaB)/\bar T$, with the galaxy fluctuation at redshift $z$ in direction $\thetaB'$, $\delta_n(\thetaB',z)=\delta n(\thetaB',z)/\bar{n}$, gives
\begin{equation}
\left< \delta_T (\thetaB) \delta_n(\thetaB',z)\right> = w_{nT}(\theta,z)
\end{equation}
where $\cos\theta=\thetaB \cdot \thetaB'$. This scale and redshift dependent signal provides information about the growth rates of large--scale structure at the redshift of the galaxy sample.

Gravitational lensing alters the observed galaxy fluctuation in two ways. First, lensing alters the area of the patch of sky being observed thus modifying the apparent number density. Second, lensing can focus light promoting intrinsically faint objects above the magnitude threshold thus increasing the measured number density \cite{N89,BTP95}. Including these effects changes the observed galaxy fluctuation to
\begin{equation}
\label{deltaneq}
\delta_n = \delta_g + \delta_{\mu}
\end{equation}
where $\delta_g$ is the intrinsic galaxy fluctuation and $\delta_\mu$ is the \emph{magnification bias} correction due to gravitational lensing. The effect of magnification bias on the galaxy--galaxy and galaxy--quasar correlation functions is well studied \cite{BTP95,V96,VFC97,MJV98,MJ98,K98}. The most recent measurements of this effect are discussed
in \cite{EGmag03,JSS03,ScrantonSDSS05}. 
Discussions of earlier measurements can be found in the references therein.
\begin{figure}[tb]
\centerline{\epsfxsize=8cm\epsffile{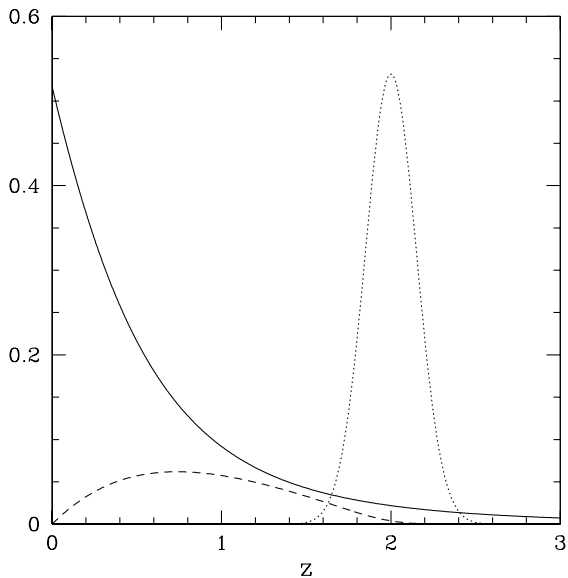}}
\caption{The derivative $(d/dz)[D(z)(1+z)]$ (solid line) where $D$ is the linear growth factor, 
a galaxy selection function (dotted line) and the corresponding lensing weight function divided by $c/H_0$ (dashed line). 
The galaxy--temperature correlation is proportional to the integral of the product of the solid and dotted lines,
while the magnification--temperature correlation is proportional to the integral of the product of the solid and dashed lines.}
\label{selraddgrow}
\end{figure}

With magnification bias the cross--correlation signal becomes 
\begin{equation}\label{wtheq}
w_{nT}(\theta,z)=w_{gT}(\theta,z)+w_{\mu T}(\theta,z).
\end{equation}
The galaxy--temperature term, $w_{gT}(\theta,z)$, is significant only when
dark energy is non--negligible
(assuming a flat universe, as is throughout this paper). Thus for high redshift galaxy samples this term is very small.
However the magnification--temperature term, $w_{\mu T}(\theta,z)$, depends on the lower--redshift distribution of lensing objects so at high redshifts it may dominate over the $w_{gT}(\theta,z)$ term.  Figure \ref{selraddgrow} illustrates this.  

We examine the effect of the magnification bias term on measurements of the ISW effect from cross--correlation.  The magnification--temperature correlation has a different scale and redshift dependence from the galaxy--temperature correlation.  We will demonstrate that the magnification term can be large though its magnitude depends on the population of galaxies under consideration.

This paper is organized as follows. 
We present the basic expressions that govern the anisotropies and correlation functions in
\S \ref{anisotropies} and \S \ref{functions}.
The ISW cross--correlation signal, especially in the presence of magnification bias, is sensitive
to the sample of objects one is cross-correlating with. We define two illustrative samples
in \S \ref{samples}. The cross--correlation signals for these
two samples, with and without magnification,  are discussed in \S \ref{signal}. In \S \ref{bias}, we investigate how erroneous
conclusions regarding dark energy would be reached if one ignores magnification bias when
interpreting ISW measurements. In \S \ref{forecasts}, 
we study how projections for the signal--to--noise and
parameter estimation might be altered by the presence of magnification bias. 
We conclude with a discussion of existing measurements in \S \ref{discuss}.
Appendix \ref{appendixA} is devoted to a technical discussion of how to estimate error
and bias. In particular, we develop a formalism for predicting the estimation--bias
for parameters of interest when certain physical effects (such as
magnification bias) are ignored in interpreting data. We keep the discussion there
quite general with a view towards possible applications other than ISW measurements.

Two comments are in order before we proceed. 
Some of the existing ISW measurements come from cross-correlating the microwave
background temperature with other diffuse backgrounds such as the X-ray background e.g. \cite{BC04}. Magnification bias, an effect that affects number counts, does not affect these measurements. This is not to say that lensing has no effect on diffuse backgrounds, gravitational lensing does have an effect through stochastic deflections. In particular, one might wonder how the lensing of the microwave background itself might impact
the observed galaxy--temperature cross--correlation. The effect appears to be small on 
the large scales where the ISW
measurements are generally considered interesting (we consider $\ell = 2 - 100$). The reader
is referred to \cite{FGC03} and references therein for more discussions.

\section{Galaxy, Temperature and Magnification Bias Anisotropies}
\label{anisotropies}

The temperature anisotropy due to the ISW effect is expressed as an integral over conformal time from $0$ to today $\eta_0$ \cite{dodelson}
\begin{equation}
\delta_T^{\rm ISW} (\thetaB)=2\int^{\eta_0}_{0}\!\!\!d\eta\, \textrm{e}^{-\tau(\eta)}\frac{\partial\phi}{\partial\eta}
\label{deltaTisw}
\end{equation}
where $\tau(\eta)$ is the optical depth between $\eta_0$ and $\eta$. For perturbations sufficiently within the horizon, the gravitational potential $\phi$
is related to the mass (or matter) fluctuation $\delta = \delta\rho/\bar\rho$ in Fourier space by \cite{dodelson}
\begin{equation}
\phi({\bf k},z)=-\frac{3}{2}\frac{H_0^2}{c^2}\Omega_m (1+z)\frac{\delta({\bf k},z)}{k^2}
\end{equation}
where $\Omega_m$ is the ratio of the matter density to the critical density today, 
$H_0$ is the Hubble constant today, $c$ is the speed of light, $z$ is
the redshift, and $k$ is the comoving wavenumber.
The mass fluctuation $\delta$ on sufficiently large scales grows according to linear theory:
$\delta \propto D(z)$, where $D(z)$ is commonly referred to as the growth factor.

We are interested in cross--correlating the temperature anisotropies, $\delta_T$, with the observed galaxy overdensity, $\delta_n$. With magnification, the measured galaxy fluctuation is a sum of two terms (eq. [\ref{deltaneq}]).  The first is the intrinsic galaxy fluctuation
\begin{equation}
\delta_g(\thetaB,z_i)= \int dz \,b(z)\,W(z,z_i) \delta(\chi(z)\thetaB,z),
\label{dg}
\end{equation} 
where $b(z)$ is an assumed scale--independent bias factor relating the galaxy overdensity to 
the mass overdensity i.e. $\delta_g =b\,\delta$, 
$W(z,z_i)$ is a normalized selection function about some mean
redshift $z_i$, and $\chi(z)$ is the comoving distance to redshift $z$.

The magnification bias term is also expressed as an integral over redshift \cite{BTP95}
\begin{eqnarray}
\delta_{\mu}(\thetaB,z_i)&=&3\Omega_m\frac{H_0^2}{c^2}(2.5s(z_i)-1)\nonumber\\
&\times& \int dz \frac{c}{H(z)}g(z,z_i)(1+z)\delta(\chi(z)\thetaB,z).
\end{eqnarray}
The lensing weight function, $g(z,z_i)$, can be thought of as
being proportional to the probability of sources around $z_i$ to be lensed by intervening matter at $z$
\cite{gands}.
\begin{equation}\label{lweight}
g(z,z_i)=\chi(z)\int_z^{\infty} \!\! dz'\frac{\chi(z')-\chi(z)}{\chi(z')}W(z',z_i)
\end{equation}
For sources at a distance $\chi$, the lensing weight function peaks at $\sim\chi/2$. The prefactor of the magnification term depends on the slope of the number count of the source galaxies. This is defined as
\begin{equation}
s\equiv \frac{d \log_{10} N(<m)}{d m}
\label{sdefine}
\end{equation}
where $m$ is the limiting magnitude of one's sample, and $N(< m)$ represents the count of
objects brighter than $m$. Magnification bias vanishes in the case that $s=0.4$. 
The values of the slope $s$, as well as the galaxy bias $b$ in eq. 
[\ref{dg}], depend on the population of galaxies under consideration.

\section{Cross--Correlation Functions}
\label{functions}

For the most part, we focus on the spherical harmonic transform of the various angular correlation
functions $w_{gT}(\theta)$, $w_{\mu T}(\theta)$ and so on. They can all be neatly described by
the following expression:
\begin{equation}
\label{cc1}
C^{AB}_\ell(z_i,z_j)=\frac{2}{\pi} \int k^2 dk P(k)\delta A_\ell (k,z_i) \delta B_\ell(k,z_j)
\end{equation}
where $P(k)$ is the matter power spectrum today as a function of the wavenumber $k$, 
and the functions $\delta A_\ell$ and $\delta B_\ell$ in the integrand can
be one of the following:
\begin{equation}
\label{cc2}
\left[\delta_g\right]_\ell(k,z_i)=b_i\int dz \, W(z,z_i) D(z)j_\ell(k\chi(z))
\end{equation}
\begin{eqnarray}
\label{cc3}
\left[\delta_\mu \right]_\ell(k,z_i)=3\Omega_m \frac{H_0^2}{c^2}(2.5 s_i-1)\qquad\qquad\qquad\quad\\
\qquad\times \int dz \frac{c}{H(z)}g(z,z_i)(1+z)D(z)j_\ell(k\chi(z))\nonumber
\end{eqnarray}
\begin{eqnarray}
\label{cc4}
\left[\delta_T^{\rm ISW} \right]_\ell(k)=3\frac{H_0^2}{c^2}\Omega_m\qquad\qquad\qquad\qquad\quad\\
\quad\times\int dz \frac{d}{dz}\left[D(z)(1+z)\right]\frac{j_\ell(k\chi(z))}{k^2}\nonumber
\end{eqnarray}

The symbol $j_{\ell}$ denotes the spherical Bessel function. Here we have elected to assume a constant galaxy--bias $b_i$ and slope $s_i$ for each redshift bin centered around 
$z_i$. In the temperature integral we have neglected the factor containing the optical depth. 
This leads to an error of at most about $3\%$ at the highest redshift we consider.

Notice that the temperature function $[\delta_T]_{\ell}(k)$ is independent of redshift, so the cross--correlation functions $C^{gT}_{\ell}$ and $C^{\mu T}_{\ell}$ are functions of only a single redshift $z_i$. This is in contrast to the other correlation functions which depend on the redshifts of the two samples being correlated, e.g. $C^{gg}_{\ell}(z_i,z_j)$, $C^{g\mu}_{\ell}(z_i,z_j)$, $C^{\mu g}_{\ell}(z_i,z_j)$ and $C^{\mu\mu}_{\ell}(z_i,z_j)$. Note also that eq. [\ref{cc1}] and eq. [\ref{cc4}] are not suitable
for computing $C^{TT}_\ell$ since that would account for only the ISW contribution to $C^{TT}_\ell$.

Examining the correlation functions, we can see that the relative magnitude of the intrinsic galaxy--temperature correlation, $C^{gT}_\ell$,  and the magnification--temperature correlation, $C^{\mu T}_\ell$, is redshift and scale dependent. The lensing efficiency increases with the redshift of the source galaxies causing the magnification bias effect to increase with redshift as well.  The scale dependence arises because each term depends on the matter distribution at different distances: the galaxy term depends on the matter distribution at the source redshifts while the magnification term depends on the distribution at the lenses. 

Additionally, the coefficients of $C^{gT}_\ell$ and $C^{\mu T}_\ell$ depend on properties of the galaxies via $b$ and $s$. The magnitude of the magnification term compared with the galaxy term depends on the ratio $(2.5s-1)/b$. Thus the correction may be negative for ($s<0.4$) and in any case increases in magnitude as $|s-0.4|$. The bias and slope depend on the choice of galaxy sample and are redshift dependent themselves. 

The Limber approximation, which is quite accurate when $\ell$ is not
too small ($\ell \gsim 10$),
can be obtained from eq. [\ref{cc1}] by setting $P(k) = P(k=(\ell+1/2)/\chi(z))$ and using the
fact that $(2/\pi)\int k^2 dk j_\ell (k\chi) j_\ell (k\chi') = (1/\chi^2) \delta (\chi-\chi')$. We find that the substitution $k=(\ell+1/2)/\chi(z)$ is a better approximation to the exact expressions than $k=\ell/\chi(z)$  (see also \cite{ALS04}).  
More explicitly, $C_\ell^{gT}$ and $C_\ell^{\mu T}$ under the Limber approximation are given by:
\begin{eqnarray}
C_\ell^{gT}(z_i) = {3\Omega_m H_0^2 \over c^2} {b_i\over ( \ell+1/2)^2} \int dz W(z,z_i) {H(z) \over c} D(z) \nonumber \\ 
\times \quad {d\over dz}[D(z)(1+z)] P(k_\perp = (\ell+1/2)/\chi) \qquad \quad\\
C_\ell^{\mu T} (z_i) = \left(3\Omega_m H_0^2 \over c^2\right)^2 {2.5 s_i - 1 \over (\ell+1/2)^2}
\int dz \,g(z,z_i)  (1+z)\nonumber \\ 
\times\quad D(z) {d\over dz}[D(z)(1+z)] 
P(k_\perp =(\ell+1/2)/\chi)\qquad \quad
\end{eqnarray}
When displaying the above quantities in figures, we follow the custom of multiplying them
by $2.725$ K. 

Unless otherwise stated, we adopt throughout this paper the following
values for the cosmological parameters when making
predictions for the various correlations of interest: we assume a flat universe with
a matter density of $\Omega_m = 0.27$, a cosmological constant of 
$\Omega_{\Lambda} = 0.73$, 
a Hubble constant of $h = 0.7$, a baryon density of $\Omega_b = 0.046$, 
a scalar spectral index of $n_s = 0.95$ and a fluctuation amplitude of $\sigma_8 = 0.8$. 
The matter power spectrum is computed using the transfer function of Eisenstein and Hu \cite{EH98}.
The microwave background temperature power spectrum $C^{TT}_\ell$ is computed
using the publicly available code CAMB \cite{CAMB}, with an optical depth of $\tau = 0.09$ and no tensor modes.
Throughout this paper, we consider ISW measurements at $\ell = 2 - 100$. 
Including higher $\ell$ modes does not significantly change our conclusions. This is
because the signal-to-noise of ISW measurements is dominated by $\ell = 10 - 100$ (see e.g. \cite{huscranton,afshordi04}, but also \cite{caveat}).
We adopt the Limber approximation for all discussions 
in \S \ref{bias} and \ref{forecasts}, which concern issues related to signal--to--noise,
estimation--bias and parameter forecast, since these issues are not significantly
affected by the very low $\ell$ modes.
The discussions and figures in \S \ref{signal}, however, make use of the exact expressions
for the correlation functions (eq. [\ref{cc1}]).

\section{Survey Samples}
\label{samples}

\begin{figure}[tb]
\centerline{\epsfxsize=9cm\epsffile{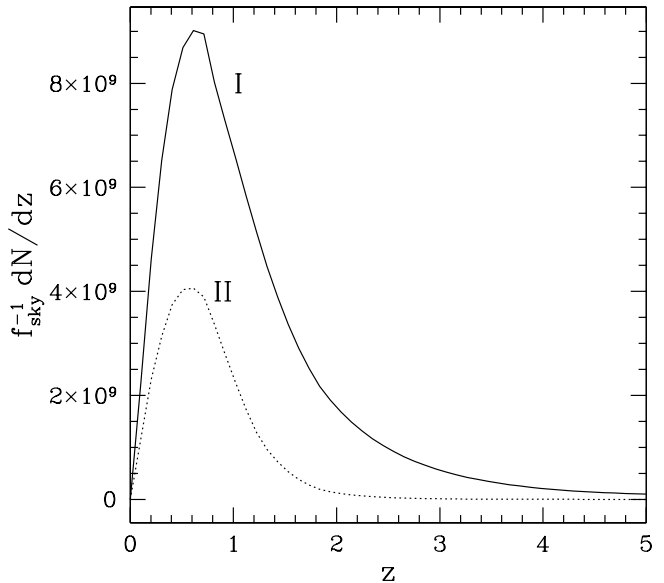}}
\caption{The number of galaxies per unit redshift over the whole sky as a function of redshift, for
two illustrative galaxy samples I and II (see text). The actual $dN/dz$ for a given survey
equals the sky coverage $f_{\rm sky}$ times the above.}
\label{dNdz}
\end{figure}

\begin{table}
\begin{small}
\begin{tabular}{|c|c|c|c|c|c|c|}
\hline \hline
 \, Sample-I & & & & & & \\
$m <27$ & z-bin 1 & z-bin 2 & z-bin 3 & z-bin 4 & z-bin 5 & z-bin 6 \\
\hline
$z_i$ & 0.49 & 1.14 & 1.93 & 2.74 & 3.54 & 4.35 \\
$b_i$ & 1.08 & 1.37 & 2.02 & 2.90 & 3.89 & 4.81 \\
$s_i$ & 0.15 & 0.20 & 0.31 & 0.43 & 0.54 & 0.63 \\
$n_i$ & 34.6 & 29.0 & 10.1 & 3.89 & 1.68 & 0.81 \\
\hline \hline
 \, Sample-II & & & & & & \\
$m <25$  & z-bin 1 & z-bin 2 & z-bin 3 & z-bin 4 & z-bin 5 & -- \\
\hline
$z_i$ & 0.48 & 1.07 & 1.85 & 2.67 & 3.46 & -- \\
$b_i$ & 1.13 & 1.51 & 2.73 & 4.57 & 6.63 & -- \\
$s_i$ & 0.19 & 0.35 & 0.86 & 1.31 & 1.75 & -- \\
$n_i$ & 16.1 & 8.6  & 0.87 & 0.11 & 0.02 & -- \\
\hline
\end{tabular}
\end{small}
\caption{The mean redshift $z_i$, 
the galaxy--bias $b_i$, the slope $s_i$ and the number of galaxies per square arcminute $n_i$ are given
for samples I and II, and for each corresponding redshift bin.}
\label{biasslope}
\end{table}

The predictions for the galaxy--temperature and magnification--temperature correlation
depend on three quantities that are sample dependent: the redshift distribution of galaxies,
the galaxy--bias, and the slope of the number count (at the magnitude cut--off). 
For illustration, we define two semi-realistic samples. 

Sample I is defined by an observed I--band (centered around $7994 \AA$)
magnitude cut of $27$ which implies a redshift distribution $dN/dz$ shown as a solid line
in Figure \ref{dNdz},
adopting the observed redshift-dependent luminosity function given by \cite{gabasch}. This gives a
net galaxy angular number density of $80$ per square arcminute. We assume a sky coverage of
$f_{sky} = 0.5$. These survey specifications are similar to those of the Large Synoptic
Survey Telescope \cite{margoniner}. We divide these galaxies into 6 redshift bins, following
the procedure of \cite{huscranton}:
\begin{eqnarray}
\label{zbin}
W(z,z_i) \propto {1\over 2} {dN(z) \over dz} 
\Bigl[{\,\rm erfc}\left({(i-1)\Delta -z \over \sigma(z)\sqrt{2}}\right) 
\\ \nonumber - {\,\rm erfc}
\left({i\Delta - z \over \sigma(z)\sqrt{2}}\right)\Bigr]
\end{eqnarray}
where $\Delta = 0.8$, $\sigma(z) = 0.02(1+z)$, and $z_i$ denotes
the mean redshift of the bin $i$. 
The complementary error function is defined as ${\,\rm erfc}(x) = (2/\sqrt{\pi})\int_x^\infty
{\,\rm exp}(-t^2) dt$. 
The normalization of $W$ is fixed by
demanding that $\int dz W(z,z_i) = 1$. 
We have checked that none of our conclusions are significantly altered by increasing
the number of bins.

Given the observed luminosity function from \cite{gabasch}, the
slope $s$ (eq. [\ref{sdefine}]) at the specified apparent magnitude cut 
can be obtained in a straightforward manner. The galaxy--bias $b$, on the other hand,
requires some theoretical input. Here, we adopt the method of Kravtsov et al. 
\cite{kravtsov}. It works basically by number matching: one matches the
the number of galaxies brighter than some magnitude with the number of
halos and subhalos above some mass threshold. Once this mass threshold is known,
the galaxy--bias can be computed. More specifically, the number of (central plus
satellite) galaxies within a halo of a given mass $M$ is
\begin{eqnarray}
{1\over 2} {\,\rm erfc}\left({{\,\rm ln}M_0 - {\,\rm ln}M \over 
0.1 \sqrt{2}}\right) \times (1 + {M \over 20 M_0}) \\ \nonumber
\end{eqnarray}
Here, the complementary
error function gives essentially a step function with a modest spread
(of $0.1$), and $M_0$ can be thought of as the mass threshold.
The quantity $20 M_0$ specifies the (parent) 
halo mass at which the expected
number of satellites is unity. This factor is in principle redshift dependent, but
we ignore such complications here.
We determine $M_0$ by demanding that the integral of the halo occupation
number over the halo mass function equals the number of galaxies for the sample
under consideration.
The bias of this sample of galaxies can be obtained by integrating the product of the halo occupation
number and the halo--bias over
the halo mass function, normalized by the total number of galaxies.
We adopt the mass function and halo-bias of Sheth and Tormen 
\cite{shethtormen}.

\begin{figure}[tb]
\centerline{\epsfxsize=9cm\epsffile{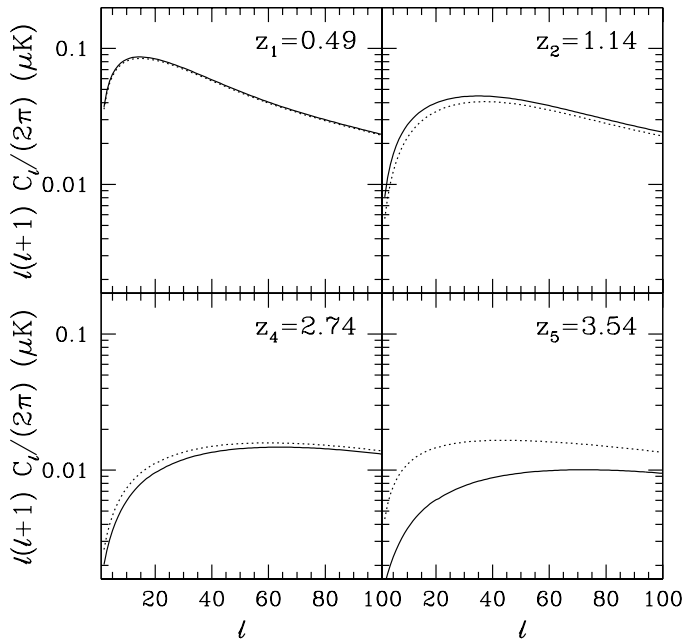}}
\caption{The galaxy--temperature correlation (solid lines) compared with the net
galaxy--temperature plus magnification--temperature correlation (dotted lines) for four
different redshift bins.
This is for sample I.}
\label{Cisw27}
\end{figure}

\begin{figure}[tb]
\centerline{\epsfxsize=9cm\epsffile{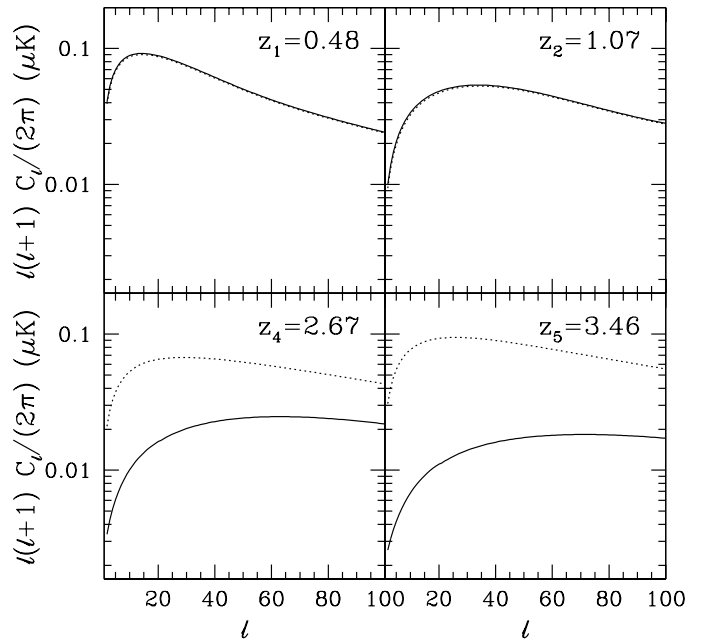}}
\caption{The analog of Figure \ref{Cisw27} for sample II.}
\label{Cisw25}
\end{figure}

Sample II is defined in a similar way to sample I, except that the
limiting I--magnitude is taken to be $25$ instead of $27$. The corresponding $dN/dz$ is
shown as a dotted line in Figure \ref{dNdz}. The net galaxy angular number density
is $26$ per square arcminute. We assume the same sky coverage of $f_{sky} = 0.5$. 
This is a shallower survey, and we divide the sample into only 5 redshift bins i.e.
apply eq. [\ref{zbin}] up to $i=5$. 

A summary of the mean redshift, the galaxy--bias $b$, the slope $s$ and the galaxy number density
for each redshift bin and each sample is given in Table \ref{biasslope}. The tabulated $dN/dz$ for each sample is available from \cite{mariweb}.
\section{The Cross--Correlation Signal}
\label{signal}

In Figures \ref{Cisw27} and \ref{Cisw25} we show the cross--correlations 
$C^{gT}_\ell (z_i)$ and $C^{nT}_\ell (z_i) = C^{gT}_\ell (z_i) + C^{\mu T}_\ell (z_i)$ at several redshift bins for both sample I and sample II, calculated using equations
(\ref{cc1}) - (\ref{cc4}).

There are several features in Figures \ref{Cisw27} and \ref{Cisw25} that
are worth commenting.
As is expected, the galaxy--temperature correlation $C^{gT}_\ell$ decreases
rapidly with redshift: at high $z$, dark energy becomes subdominant
making the time derivative of the gravitational potential quite small
(eq. [\ref{deltaTisw}]). Note that this decrease of $C^{gT}_\ell$ occurs
despite the overall increase of the galaxy--bias with redshift. 

The magnification--temperature correlation $C^{\mu T}_\ell$, on the other hand,
generally increases with redshift because of the increase of the lensing
efficiency. The magnification--temperature correlation $C^{\mu T}$ can
be high even as $C^{gT}$ becomes small, because the rather broad lensing weight function
makes $C^{\mu T}$ sensitive to structures at low redshifts where the gravitational potential
has a significant time derivative.
The general increase of $C^{\mu T}_\ell$ with
redshift is aided also by the increase of the number count slope
$s$ --- at a higher $z$ one is generally looking at intrinsically brighter
galaxies which reside in the steep part of the luminosity function.
The net result is that the total cross-correlation $C^{nT}_\ell$ starts to climb
with redshift, for $z \gsim 2$. 
In fact at a sufficiently high $z$, $C^{nT}_{\ell}$ even 
becomes comparable with $C^{nT}_\ell$ or $C^{gT}_\ell$ at the lowest redshift bin.
Comparing sample I and sample II, one can see that a brighter magnitude cut
leads to a more pronounced magnification bias effect. This is due to the steeper number count slope.

Another feature visible in Figures \ref{Cisw27} and \ref{Cisw25} is that magnification changes the shape of $C^{nT}_{\ell}$ with $\ell$. The magnification term, being sensitive to structures at lower redshift, peaks at a lower $\ell$ than the galaxy term. For positive magnification contribution, the peak of $C^{nT}_{\ell}$ is at a lower $\ell$ than the peak of $C^{gT}_{\ell}$. 

\begin{figure}[tb]
\centerline{\epsfxsize=8cm\epsffile{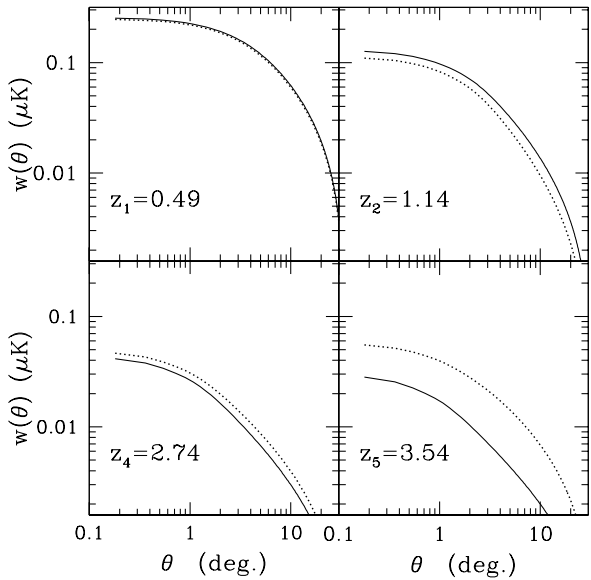}}
\caption{The 2-point galaxy--temperature correlation $w_{gT}(\theta,z_i)$ (solid lines) compared with the net galaxy--temperature plus magnification--temperature correlation $w_{nT}(\theta,z_i)$ (dotted lines) in several of our redshift bins. This is for sample I.}
\label{wth27fig}
\end{figure}

\begin{figure}[tb]
\centerline{\epsfxsize=8cm\epsffile{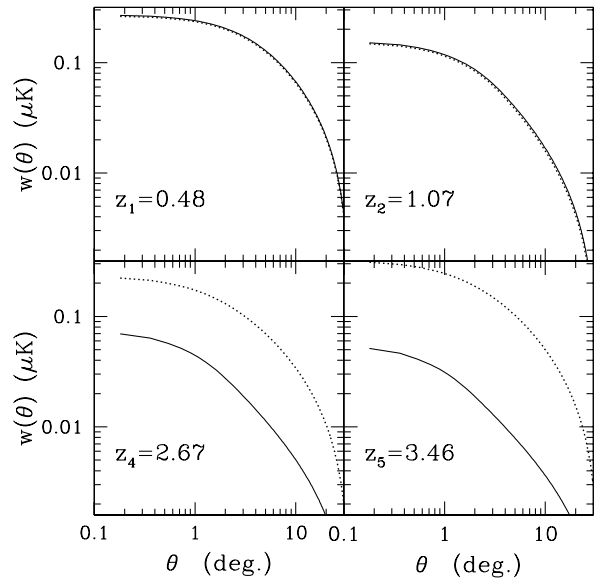}}
\caption{The analog of Figure \ref{wth27fig} for sample II.}
\label{wth25fig}
\end{figure}

In Figures \ref{wth27fig} and \ref{wth25fig} we plot the real space correlation functions $w_{gT}(\theta,z_i)$ and $w_{nT}(\theta,z_i)$ without the monopole and dipole contributions. These are calculated from
\begin{equation}
w_{X}(\theta,z_i)=\sum_{\ell=2}^{500} \frac{2 \ell+1}{4\pi}C^{X}_\ell(z_i) P_{\ell}(\cos\theta).
\end{equation}
where $X$ symbolizes $gT$ or $nT$ and $P_{\ell}(\cos\theta)$ are the Legendre polynomials.  We have extended the range of the sum to $\ell=500$ to insure the convergence of the sum for small values of $\theta$.  The general discussion about $C^{nT}_{\ell}$ applies to $w_{nT}$ as well; at $z\gsim 2$ magnification bias grows to become an important component of the cross--correlation signal, and the amplitude is preferentially increased at large angular scales. 

Magnification bias can of course have either sign. 
At low redshifts, one is generally looking at intrinsically faint galaxies
which leads to a small $s$ that can be less than $0.4$. This is why 
$C^{nT}_\ell < C^{gT}_\ell$ at low redshifts for our samples.
The transition redshift from $s < 0.4$ to $s > 0.4$ is sample dependent.
For a sample with an apparent limiting magnitude cut, the brighter the cut,
the lower the transition redshift. 
It is conceivable that a sufficiently
faint cut can be achieved such that $s$ remains less than $0.4$ out to
redshifts where $|C^{\mu T}_\ell| > C^{gT}_\ell$. In such a case
the net cross-correlation $C^{nT}_\ell$ would become negative. 
This, however, requires an exceptionally deep survey: an I-band magnitude limit of 28 or higher,
according to the observed luminosity function \cite{gabasch}.
Conversely, if a sufficiently bright galaxy cut were used at low redshift it is possible that magnification 
could become important at lower redshifts than evidenced by our samples. 

In the literature the cross--correlation signal is often presented in the normalized form $C^{gT}_{\ell}(z_i)/\sqrt{C^{gg}_{\ell}(z_i)}$ or $w_{gT}(\theta,z_i)/\sqrt{w_{gg}(\theta,z_i)}$ to remove the dependence on galaxy bias. However, when magnification bias is taken into account the auto--correlation is modified to be
\begin{eqnarray}\label{autocorr}
C^{nn}_{\ell}(z_i,z_j) = \qquad \qquad\qquad\qquad\qquad\qquad\qquad\qquad\qquad \\ \quad C^{gg}_{\ell}(z_i,z_j)+C^{g\mu}_{\ell}(z_i,z_j)+C^{\mu g}_{\ell}(z_i,z_j)+C^{\mu\mu}_{\ell}(z_i,z_j)\nonumber .
\end{eqnarray}
Note, then, that the galaxy--bias would be completely removed by the division only if magnification bias were absent. 
Had we plotted the normalized versions of the cross-correlation signals, the fractional difference
between $C^{gT}_\ell/\sqrt{C^{gg}_\ell}$ and $C^{nT}_\ell/\sqrt{C^{nn}_\ell}$ would remain about the same as
that between the un-normalized versions. This is because
for the cases considered here $C^{nn}_\ell (z_i, z_i) \sim C^{gg}_\ell (z_i, z_i)$ to within about $20 \%$.

\section{Estimation--bias due to Ignoring Magnification ---- a Thought Experiment}
\label{bias}

\begin{figure}[tb]
\centerline{\epsfxsize=8cm\epsffile{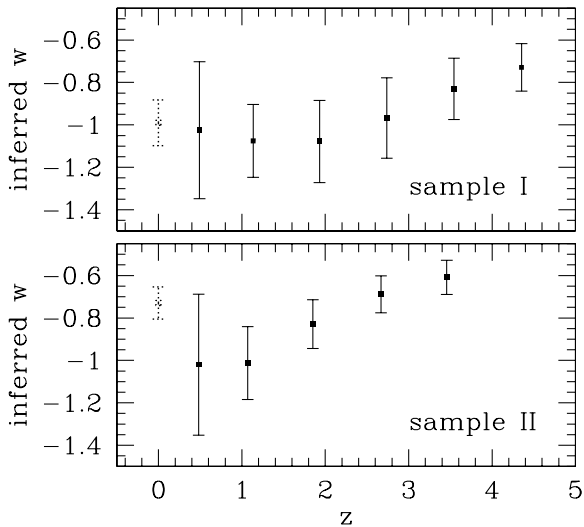}}
\caption{The average 
{\it inferred} dark energy equation of state $w$ from ISW measurements when (erroneously) ignoring magnification.
The true $w$ equals $-1$.
The solid symbols show the inferred $w$ for each redshift bin separately. 
The dotted symbols at $z = 0$ show the inferred $w$ combining all redshift bins. 
The error-bars are 1 $\sigma$ --- note that they are dependent upon the {\it inferred} model (see text).
The upper panel is for sample I, and the lower panel is for sample II.}
\label{wbiasfig}
\end{figure}

Clearly, magnification bias should be taken into account when interpreting ISW measurements, especially at high redshifts
where the magnification--temperature correlation actually dominates over the usual galaxy--temperature signal.
Ignoring it would lead to erroneous conclusions about dark energy. 
We quantify this by computing the estimation--bias in the dark energy equation of state $w$ 
if magnification were ignored when interpreting measurements from samples I and II.

A thought experiment is set up as follows. Suppose the true dark energy equation of state is $w = -1$ i.e.
a cosmological constant (and other cosmological parameters take the values stated at the
end of \S \ref{functions}). Suppose further one were to infer $w$ (and other parameters) 
from measurements of the total cross-correlation
$C^{nT}_\ell$ by fitting them with a model that ignores magnification i.e. $C^{gT}_\ell$, and
one were to assign errorbars based on the (erroneous) model with no magnification.
The fitted parameters include: the dark energy equation of state $w$ (i.e. $w = P/\rho$ where 
$P$ and $\rho$ are the dark energy pressure and energy density respectively), the matter
density $\Omega_m$ (we assume a flat universe so that the dark energy density is $1 - \Omega_m$),
the amplitude of fluctuations $\sigma_8$, the spectral index $n_s$,
the Hubble constant $h$, the baryon density $\Omega_b$ and the galaxy-bias for
each redshift bin $b_i$. Constraints from other data (such as from the temperature
anisotropy itself $C^{TT}_\ell$ and from the galaxy power spectrum) are simulated
by the inclusion of priors: the fractional (1 $\sigma$) errors on 
$\Omega_m h^2$, $\Omega_b h^2$, $h$ and $\sigma_8$
are assumed to be $5 \%$, the fractional error on $b_i$ is $10 \%$, and $n_s$ has an (absolute) error of
$0.02$. These are similar to current constraints, depending somewhat on assumptions \cite{maxlrg}.
A simple analytic expression can be derived for the parameter estimation--bias, which is discussed in
detail in Appendix A (eq. [\ref{wbias}]).

The inferred $w$, together with its marginalized error, for samples I and II are shown in
Figure \ref{wbiasfig}. The modes used range from $\ell = 2$ to $\ell = 100$.
The solid symbols show the inferred $w$ from each redshift bin separately. 
The sign of the resulting estimation--bias is determined by the slope $s$. 
If $s < 0.4$, as is the case at low redshifts, the inferred $w$ tends to be lower
than the true value. This is understandable because the {\it observed} $C^{nT}$ is
suppressed by magnification in that case, and the only way to stick with an
erroneous model without magnification is to
resort to a lower $w$ to delay dark energy domination.
The reverse happens at high redshifts, where $s > 0.4$. In fact, at the highest 
redshift bin, the bias can become quite large. Ignoring
magnification bias is simply unacceptable.

It is worth emphasizing that the errorbars are model dependent i.e. they
are computed using the {\it inferred} model, which is different
for each redshift bin. The general trend is for the errorbar to shrink as
one goes to higher redshifts, despite the drop in signal--to--noise (see \S \ref{forecasts}).
This is due to the fact that $d{\,\rm ln} C^{gT}/dw$ rises with both $z$ and
$w$. The generally positive bias in $w$ at high redshifts helps to diminish the
corresponding errorbars.

The dotted symbols at $z = 0$ show the inferred $w$ and its associated errorbar
when measurements from all the redshift bins are combined.
The previous remarks on the model dependent nature of the errorbar applies here as well.
In the case of sample II, one finds that the overall inferred $w$ is biased high by
almost 3 $\sigma$.

\section{Signal--to--Noise and Parameter Forecast}
\label{forecasts}

\begin{figure}[tb]
\centerline{\epsfxsize=9cm\epsffile{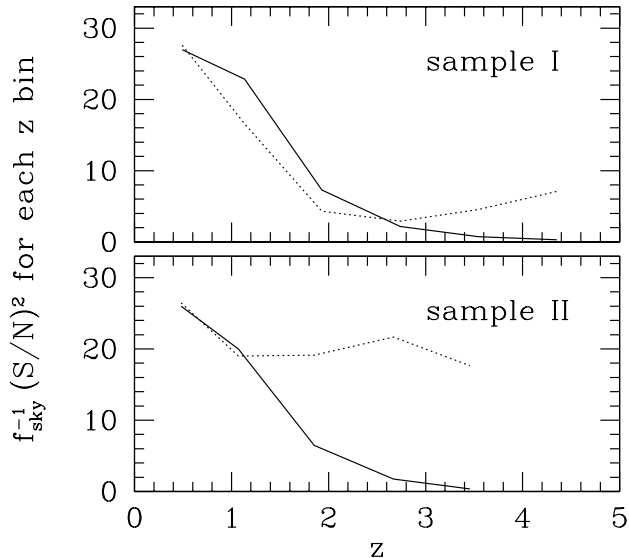}}
\caption{The signal--to--noise (squared) of ISW measurements from each redshift bin for samples I and II.
The solid line ignores magnification (equivalent to setting $s = 0.4$) while the dotted
line includes magnification. Note $(S/N)^2 \propto f_{\rm sky}$ i.e. the division by $f_{\rm sky}$
takes out the dependence on sky coverage.}
\label{dSN}
\end{figure}

\begin{figure}[tb]
\centerline{\epsfxsize=9cm\epsffile{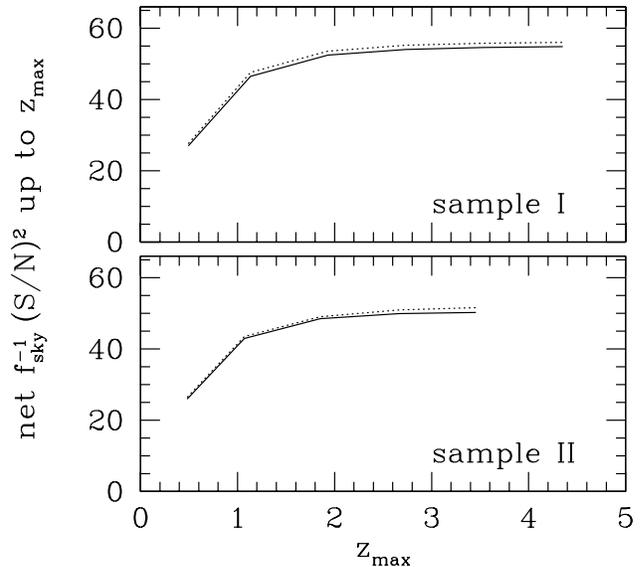}}
\caption{The net cumulative 
signal--to--noise (squared) of ISW measurements from all redshift bins up to $z_{\rm max}$, for samples I and II.
The solid line ignores magnification (equivalent to setting $s = 0.4$) while the dotted
line includes magnification (the two are very close to each other). 
The division by $f_{\rm sky}$ removes the dependence on sky coverage.}
\label{SN}
\end{figure}

The above considerations suggest that magnification could cause
the signal--to--noise of ISW measurements to remain favorable even at relatively
high redshifts, deep into the matter dominated regime.
This is borne out by Figure \ref{dSN} \cite{comments}, which shows (with $f_{\rm sky}$ divided out)
\begin{eqnarray}
\label{binSN}
\left[{S\over N}(z_i)\right]^2 = \sum_\ell {[C^{nT}_\ell(z_i)]^2 f_{\rm sky} (2\ell + 1) \over
(C^{nn}_\ell(z_i,z_i) + {1 \over n_i}) C^{TT}_\ell + [C^{nT}_\ell (z_i)]^2}
\end{eqnarray}
where $n_i$ is the mean surface density of galaxies, and we include modes
from $\ell = 2$ to $\ell = 100$ (as in the rest of the paper). The result for sample II is particularly
striking: the signal--to--noise remains more or less flat out to high redshifts
(dotted line), in sharp contrast with the signal--to--noise if magnification is ignored
(solid line; equivalent to putting $s = 0.4$, or replacing
the superscript $n$ by $g$ in eq. [\ref{binSN}]). 

Magnification bias, therefore, opens up a high redshift window for ISW measurements which
otherwise would not exist. That is the good news. The bad news, however, is that
since the magnification of redshift $z_i$ galaxies probes structures at redshifts $z < z_i$,
the $C^{nT}_\ell$ measurements at high redshifts are in fact quite correlated with
those at low redshifts. Taking into account such correlations, we show in Figure \ref{SN} the
net accumulated signal--to--noise (squared) for measurements from all redshift bins up to a given $z_{\rm max}$,
defined as follows:
\begin{equation}
\label{sneq}
\left[\frac{S}{N}(z_{\rm max})\right]^2=\sum_{z_i,z_j \le z_{\rm max}} \sum_{\ell}C^{nT}_{\ell}(z_i)\left[\textrm{Cov}_{\ell}^{-1}\right]_{ij}C^{nT}_{\ell}(z_j)
\end{equation}
where 
\begin{eqnarray}
\label{covmat}
\left[\textrm{Cov} _{\ell}\right] _{ij}=\qquad\qquad\qquad\qquad\qquad\qquad\qquad\qquad\qquad\\ \frac{(C^{nn}_{\ell}(z_i,z_j)+\delta_{ij}/n_i)C^{TT}_{\ell}+C^{nT}_{\ell}(z_i)C^{nT}_{\ell}(z_j)}{f_{\rm sky}(2\ell+1)}.\nonumber
\end{eqnarray}

It appears the cumulative signal--to--noise reaches a plateau by $z \sim 2$ whether or not magnification bias is included. As is expected, magnification bias makes much less of a difference in this case: the solid and dotted lines are very close together. 
This happens despite the significant difference in signal--to--noise on a redshift bin by redshift bin basis, shown in Figure \ref{dSN}. The culprit is the significant correlation between different redshift bins introduced by lensing --- the high redshift measurements are in fact only probing large--scale structure at low redshifts. 

The reader might also wonder why the two samples have such similar cumulative signal--to--noise.
This is because at low redshifts (where most of the cumulative signal--to--noise comes from), 
neither sample is shot-noise dominated i.e.
for the most part, the noise is dominated by the term associated with $C^{nn}_\ell C^{TT}_\ell$
(eq. [\ref{covmat}]).

\begin{figure}[tb]
\centerline{\epsfxsize=9cm\epsffile{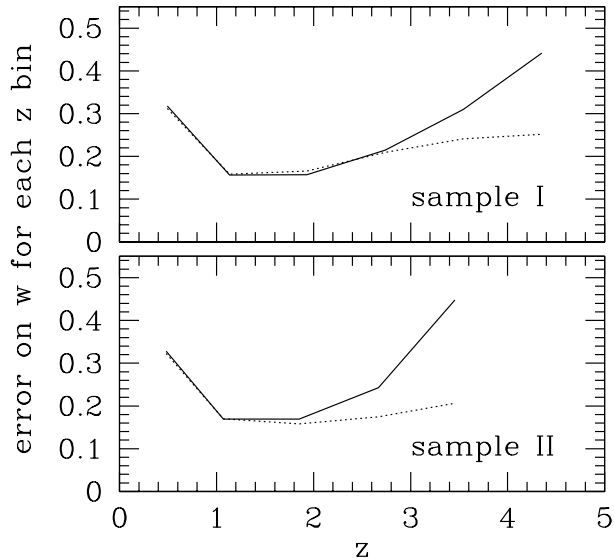}}
\caption{The marginalized error on $w$, $\sqrt{\langle \delta w^2 \rangle}$, for ISW measurements from each redshift bin. 
The solid line ignores magnification (equivalent to setting $s = 0.4$) while the dotted
line includes magnification.}
\label{dwerr}
\end{figure}

\begin{figure}[tb]
\centerline{\epsfxsize=9cm\epsffile{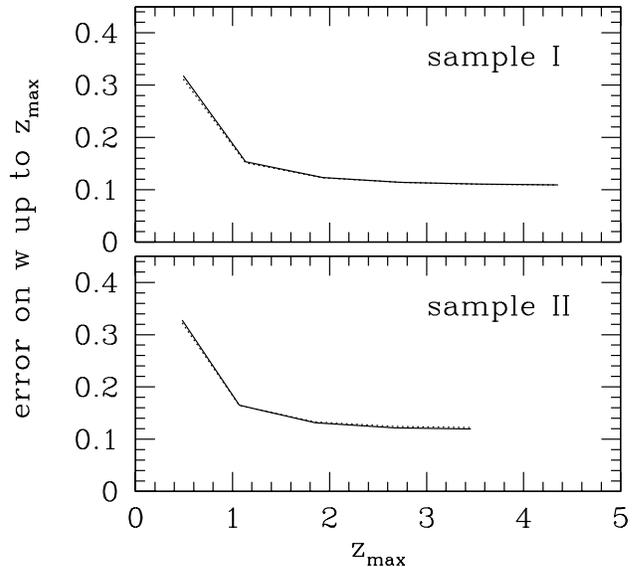}}
\caption{The marginalized error on $w$, $\sqrt{\langle \delta w^2 \rangle}$, for ISW measurements from all redshift bins combined
up to $z_{\rm max}$. The solid line ignores magnification (equivalent to setting $s = 0.4$) while the dotted
line includes magnification (the two are very close together).}
\label{werr}
\end{figure}

Ultimately, we are interested in cosmological constraints from ISW measurements on, for instance, the
dark energy equation of state $w$.
It is therefore useful to show the expected errors for $w$, both
on a redshift bin by redshift bin basis (as in Figure \ref{dSN}) and on a cumulative basis (as in Figure \ref{SN}) (see also \cite{afshordi04}). This is displayed in Figures \ref{dwerr} and \ref{werr}. 

The fiducial cosmological model here, as is in the case of the computation of $S/N$, is that described at the end of 
\S \ref{functions}. The cosmological parameters that are marginalized over and the
set of priors assumed are identical to those described in \S \ref{bias} (with the addition of
$s_i$, the number count slope for each redshift bin, as a new parameter, for which we impose
a prior of $\Delta s_i = 0.02$), and the Fisher matrix
formalism for doing so is explained in Appendix A, in particular eq. [\ref{Fprimen}].
Once again, we see that while the presence of magnification bias improves
the errors at high redshifts on a redshift bin by redshift bin basis, it does not 
in fact add much cumulative information if one considers the errors from combining
redshift bins. We have experimented with changing the priors and varying sample definitions, 
and it appears this conclusion is quite robust.

The most compelling reason for doing high redshift ($z \gsim 2$) ISW measurements is therefore
not that the expected cumulative error on $w$ would continue to improve, but that
there might be surprises e.g. dark energy might be quite different from the cosmological constant and
actually remain a significant component of the universe even at high redshifts, or gravity may be modified in non-trivial ways on large scales \cite{Zhang06, GPV04, P05,PCSCN05,modgrav}.
At the very least, high redshift ISW measurements constitute a consistency check that one should make.
As explained in \S \ref{bias}, such high redshift measurements should be interpreted with care:
magnification bias must be taken into account.

\section{Discussion}
\label{discuss}

\begin{figure}[tb]
\centerline{\epsfxsize=8cm\epsffile{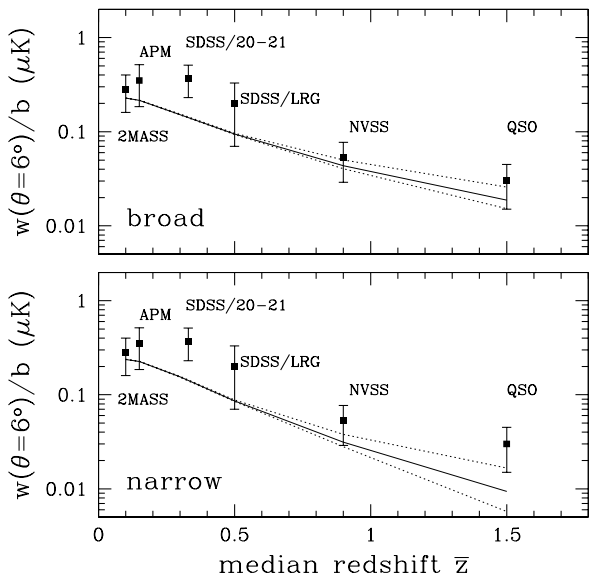}}
\caption{The two panels show the 2-point correlation function divided by the galaxy bias, $w(\theta,\bar{z})/b(\bar{z})$, 
at $\theta=6^o$ as a function of the median redshift of the sources, $\bar{z}$.  
The points with errorbars are the existing measurements. The solid line shows the
theoretical prediction ignoring magnification for a flat cosmological constant model (see \S \ref{functions}).
The dotted lines include 
magnification for the cases $s=0.8$ (upper dotted line) and $s=0.2$ (lower dotted line). 
The upper panel uses a broad galaxy selection function while the lower panel uses a narrow selection function (see text).}
\label{wtgData}
\end{figure}

A natural question: to what extent should we worry about
magnification bias when interpreting existing ISW measurements? 
This is partially addressed in Figure \ref{wtgData}, which shows the cross-correlation signal
at $6^o$ separation. 

The points with errorbars are measurements taken from a compilation in
\cite{Gaztanaga05} with the addition of recent results from \cite{CGMFS06} and \cite{QSOISW}. 
All involve correlating the microwave background with galaxies, except the highest redshift one \cite{QSOISW}
which involves correlation with quasars. The median redshifts of the measurements are: 
$\bar z \sim 0.1, 0.15, 0.3, 0.5, 0.9, 1.5$. The ISW signal shown is normalized by the galaxy/quasar bias,
with the bias values taken from \cite{CGMFS06,QSOISW,Gaztanaga05}. 
They are, in the order of increasing redshift: $1.1, 1.0, 1.2, 1.4, 1.7, 2.3$.

The solid lines show the theoretical prediction for $w_{gT}/b$ (i.e. ignoring magnification)
adopting the cosmological parameters stated in \S \ref{functions}.
Because the selection function is somewhat uncertain, we choose to illustrate the prediction with
a `broad' and a `narrow' selection function, parametrized as follows:
\begin{equation}
W(z)=\frac{\beta}{\Gamma\left[\frac{\alpha+1}{\beta}\right]} \frac{z^{\alpha}}{z_0^{\alpha+1}}e^{-(z/z_0)^\beta}.
\end{equation}
where $\alpha=2$, $\beta=1.5$ is broad, and $\alpha=2\,(1+3\,z_0)$, $\beta=\alpha$ is narrow.
For each selection function $z_0$ is adjusted to reproduce the median redshifts of the measurements.

The dotted lines \cite{lines} in Figure \ref{wtgData} show the theoretical prediction for $w_{nT}/b$ (i.e. accounting for magnification)
for $s=0.8$ (upper) and $s=0.2$ (lower). These values for the number count slope seem to
span the range observed in quasar samples \cite{ScrantonSDSS05} 
similar to the one used in the highest redshift ISW measurement, where magnification bias is most relevant.

Considering all angular scales and both selection functions, we find that for the four lowest redshift measurements 
$|w_{\mu T}/w_{gT}|< 0.1$, unless $s > 1.4$ or $s< -0.6$. For the NVSS sample at $\bar{z}\sim 0.9$, we estimate the slope of 
the NVSS radio sample to be $s=0.320 \pm 0.07$ which results in $|w_{\mu T}/w_{gT}| \lsim 0.1$. This is small compared with the 
measurement error. It appears that the quasar--temperature correlation at $\bar{z}\sim1.5$ is the only measurement for which 
magnification \emph{could} be significant. To determine precisely to what extent this measurement is affected by magnification 
requires a more detailed analysis of the quasar sample. It is conceivable that a suitably chosen quasar subsample could
exhibit a large magnification effect.

Figure \ref{wtgData} also shows that for $z\gsim1$ there is a significant dependence on the width of the selection function. 
This is because the mass--temperature correlation decreases rapidly with redshift. If a broad selection function is used, 
the galaxy--temperature correlation will decline less rapidly because low redshift sources will contribute. 
On the other hand, $w_{\mu T}$ depends on the lensing weight function (eq. [\ref{lweight}]) 
which is broadly distributed regardless of the width of the source distribution. 
In other words, changing the width of the selection function from broad to narrow causes the galaxy--temperature term to 
decrease, while leaving the magnification--temperature term largely unchanged. 
Figure \ref{wtgData} illustrates this for a fixed $\theta$, however, the statement holds for other angles.
Source distributions used in the existing ISW measurements are probably closer to the broad selection function.

To conclude, we find that magnification bias alters the observed galaxy/quasar--temperature cross--correlation significantly
at high redshifts (Figures \ref{Cisw27} and \ref{Cisw25}). 
The precise magnitude of the modification depends on the sample of objects
under consideration. Three facts more or less guarantee magnification plays a non--negligible or even dominant
role at $z \gsim 2$: the generally steepening slope of the number count at high redshifts (because
one is looking at intrinsically brighter objects), 
the decay of the intrinsic galaxy--temperature correlation $C^{gT}$
as one enters the matter--dominated era, and the increase of the lensing efficiency with redshift.
Ignoring magnification bias when interpreting high redshift ISW measurements would lead to erroneous conclusions about
the nature of dark energy. For instance, the estimated equation of state $w$ can differ
from the true one by more than 3 $\sigma$ in some cases (Figure \ref{wbiasfig}). 

The boosting of the ISW signal by magnification (assuming $s > 0.4$ which is generally the case
at high redshifts when one is looking at intrinsically bright objects) implies that, despite naive expectations, ISW measurements remain viable even at $z \gsim 2$.
However, because of the correlated nature of the lensing signal (across different redshifts), 
the cumulative information from low $z$'s to high $z$'s is not significantly
enhanced by magnification.
The most compelling scientific justification for pursuing high redshift ISW measurements
is to look for surprises: for instance, dark energy might remain significant
out to high redshifts. At the very least, one would like to perform a consistency check
of the very successful cosmological constant model of dark energy.
The large--scale nature of the ISW signal also means it is a natural place to
look for spatial fluctuations in dark energy \cite{huscranton} and possible signatures of modified 
gravity \cite{modgrav}. 

When interpreting future high redshift ISW measurements, accounting for the
effect of magnification bias is a must. The net cross--correlation signal depends not only on the cosmological parameters 
(e.g. $\Omega_m$, $\Omega_\Lambda$, $h$, $n_s$, $\sigma_8$) but also on the sample--dependent parameters $b$ and $s$. 
The slope $s$ can be estimated from the observed number counts. The bias $b$ can be inferred
by comparing the amplitude of the galaxy/quasar auto--correlation with the 
amplitude of the mass auto--correlation predicted by the microwave background, 
modulo the fact that the observed galaxy/quasar auto--correlation function is itself affected by
magnification bias (though it appears to be at most a $20 \%$ effect at the scales and redshifts
we study). Alternatively, the bias can be estimated by examining the galaxy/quasar higher order correlations.
The upshot is that separating the galaxy--temperature and the magnification--temperature signals is
in principle possible, at least in a model dependent manner. Other possibilities are to exploit the sample--dependent
nature of $b$ and $s$ to study the variation of the total ISW signal with different sample cuts, or to use the galaxy--galaxy correlation across different redshift bins to check for consistency with the ISW--magnification bias signal. 

An interesting question remains to be explored: given a survey with many different kinds 
of objects, such as galaxies/quasars of different luminosities and types, how should one go
about choosing a weighting scheme that optimizes the total signal and/or its magnification component?
The magnification component might be particularly interesting if the galaxy/quasar bias turns out to be uncertain.
We leave this for future investigation.

\acknowledgments
We would like to thank Ryan Scranton for useful discussions regarding magnification bias, Asantha 
Cooray for addressing questions we had about his results on the conditional luminosity function,
Elena Zucca for kindly sending us her group's results on the 
luminosity function for several galaxy types (see \cite{elenazucca}), and Ben Johnson for helpful discussions regarding luminosity functions and galaxy samples.
Support for this work is provided in part by the DOE, grant number DE-FG02-92-ER40699,
and the Initiatives in Science and Engineering Program at Columbia University.
EG acknowledges support from Spanish Ministerio de Ciencia y
Tecnologia (MEC), project AYA2006-06341 with EC-FEDER funding,
research project 2005SGR00728 from  Generalitat de Catalunya.

\appendix

\section{On Estimation Bias and Error}
\label{appendixA}

The technology for making error forecasts is well documented in the literature \cite{dodelson}.
What seems to be less often discussed is the issue of predicting the bias: what
is the resulting bias in the estimation of a parameter if some particular physical
effect is ignored or some approximation is made in modeling the data? 
In our case, the parameter of interest is $w$ and the physical effect
is the magnification bias. We keep our discussion here fairly general since it might
be of interest to researchers working on other problems. Much of our discussion is 
a straightforward generalization of the classic paper by Rybicki and Press \cite{rybickipress}.

In general, we have the problem of trying to obtain constraints on $m$ parameters, labeled 
$p_\alpha$ with $\alpha = 1,2, ... ,m$, from an $n$-dimensional data vector, $\hat d_i$ with $i = 1,2,...,n$.
Assuming Gaussian distributed data, the likelihood is proportional to ${\rm exp\,}[{-\chi^2/2}]$, where
\begin{eqnarray}
\label{chi2}
\chi^2 = \sum_{i,j} (\hat d_i - d_i) C^{-1}_{ij} (\hat d_j - d_j)
\end{eqnarray}
where $C_{ij} \equiv \langle (\hat d_i - d_i) (\hat d_j - d_j) \rangle$, 
with $d_i \equiv \langle \hat d_i \rangle$. 
The vector (component) $d_i$ is in general some complicated function of the parameters
$p_\alpha$. The trick is to turn this into a linear estimation problem
by supposing that $p_\alpha$ is close to some fiducial value $\bar p_\alpha$, and
therefore:
\begin{eqnarray}
\label{linearmodel}
d_i = \bar d_i + {\partial d_i \over \partial p_\alpha} \delta p_\alpha 
\quad , \quad \delta p_\alpha \equiv p_\alpha - \bar p_\alpha
\end{eqnarray}
where $\bar d_i$ is the expected data average if the parameter values were indeed
$\bar p_\alpha$, and the partial derivative is evaluated at $p_\alpha = \bar p_\alpha$.

Substituting the linear model eq. [\ref{linearmodel}] into 
the expression for $\chi^2$, we obtain
\begin{eqnarray}
\label{chi2F}
\chi^2 = \sum_{\alpha,\beta} (\delta p_\alpha - \delta \hat p_\alpha^{\rm max}) F_{\alpha\beta}
(\delta p_\beta - \delta \hat p_\beta^{\rm max}) 
\end{eqnarray}
up to an additive constant independent of $\delta p_\alpha$.
Here, 
\begin{eqnarray}
\label{dpmax}
\delta \hat p_\alpha^{\rm max} \equiv 
\sum_{\beta,i,j} 
F^{-1}_{\alpha\beta} {\partial d_i \over \partial p_\beta} C^{-1}_{ij} (\hat d_j - \bar d_j)
\end{eqnarray}
where the $\hat{}$ on top of $\delta \hat p_\alpha^{\rm max}$ is there to
remind us that this quantity depends on the data $\hat d_j$ (i.e. 
a stochastic quantity),
and $F_{\alpha\beta}$ is the Fisher matrix
\begin{eqnarray}
F_{\alpha\beta} \equiv \sum_{i,j} {\partial d_i \over \partial p_\alpha} C^{-1}_{ij} 
{\partial d_j \over \partial p_\beta}
\end{eqnarray}
Therefore, the maximum likelihood estimate for $p_\alpha$ is
\begin{eqnarray}
\label{pmax}
\hat p_\alpha^{\rm max} = \bar p_\alpha + \delta \hat p_\alpha^{\rm max}
\end{eqnarray}
and the error covariance associated with such an estimate is given by the inverse
of the Fisher matrix:
\begin{eqnarray}
\label{pperror}
\langle \hat p_\alpha^{\rm max} \hat p_\beta^{\rm max}\rangle - \langle \hat p_\alpha^{\rm max} \rangle
\langle \hat p_\beta^{\rm max} \rangle 
= F^{-1}_{\alpha\beta}
\end{eqnarray}

Combining eq. [\ref{dpmax}] and [\ref{pmax}], we see that
\begin{eqnarray}
\label{pmaxave}
&& \langle \hat p_\alpha^{\rm max} \rangle 
= \bar p_\alpha + \sum_{j} w_{\alpha j}
(\langle \hat d_j \rangle - \bar d_j) \\ \nonumber
&& w_{\alpha j} \equiv \sum_{\beta,i} F^{-1}_{\alpha\beta} {\partial d_i \over \partial p_\beta} C^{-1}_{ij} 
\end{eqnarray}
The estimation bias of interest is therefore 
$\langle \hat p_\alpha^{\rm max} \rangle - p_\alpha^{\rm true}$,
where $p_\alpha^{\rm true}$ is the true value of 
the underlying parameter. 
Note that both the weights $w_{\alpha j}$ and the vector
$\bar d_j$ are model dependent. For instance, one could choose to
ignore certain physical effects, such as magnification bias,
in constructing these quantities. The result would be a biased
estimate of $p_\alpha$. 

Let us suppose first that one is using the correct or {\it exact} model.
Then, substituting $\langle \hat d_j \rangle = \bar d_j + [{\partial d_j/\partial p_\gamma}]
(p_\gamma^{\rm true} - \bar p_\gamma)$ into eq. [\ref{pmaxave}], we have
$\langle \hat p_\alpha^{\rm max} \rangle = p_\alpha^{\rm true}$, as expected.

Suppose instead one is using an incorrect or {\it approximate} model, where a certain
physical effect (e.g. magnification bias) has been ignored. 
This means $\bar d_j$ and the weights $w_{\alpha j}$ are
all computed by ignoring this effect, whereas $\langle \hat d_j \rangle$,
which is the expectation value of the observed data, of course
has such a physical effect (and all other relevant physical effects) 
taken into account. The preceding
argument (concerning the {\it exact} model) 
no longer works and one is left with an estimation bias i.e. 
$\langle \hat p_\alpha^{\rm max} \rangle \ne p_\alpha^{\rm true}$.
Note that this bias depends on the assumed fiducial parameter
value $\bar p_\alpha$ (as opposed to the case of unbiased estimation, where
$\langle \hat p_\alpha^{\rm max} \rangle$ does not depend on $\bar p_\alpha$ at all). 
Generally, a modeler would choose
a fiducial value that is close to the {\it perceived} (i.e. possibly incorrect, due
to the use of an incorrect/approximate model) maximum likelihood value.
We therefore obtain the estimation bias by iteration: starting with
some fiducial parameter value, compute $\langle \hat p_\alpha^{\rm max} \rangle$, 
then adopt the result as a new fiducial value and iterate.
In all cases investigated in this paper, a few iterations are sufficient to guarantee convergence.
The modeler would assign a (theoretical) errorbar to his/her biased estimate
based on eq. [\ref{pperror}], where the Fisher matrix is, again, that of the
{\it approximate} model. 

Lastly, let us address the question of priors. Priors can be thought of
as just another kind of data, and they add to the $\chi^2$ in eq. [\ref{chi2}]
a term of the form:
\begin{eqnarray}
\sum_{\alpha,\beta} (\hat p_\alpha - p_\alpha) M^{-1}_{\alpha\beta} (\hat p_\beta - p_\beta) \nonumber
\end{eqnarray}
where $\hat p_\alpha$ denotes the assumed prior value (e.g. $\Omega_m = 0.27$, etc) and
$M_{\alpha\beta}$ is the covariance associated with this set of priors.
Completing square as before, one obtains, up to an irrelevant additive constant,
a $\chi^2$ slightly different from the one in eq. [\ref{chi2F}]:
\begin{eqnarray}
\chi^2 = \sum_{\alpha,\beta} (\delta p_\alpha - \delta \hat p_\alpha^{\rm max}) F'_{\alpha\beta}
(\delta p_\beta - \delta \hat p_\beta^{\rm max}) 
\end{eqnarray}
where $F'_{\alpha\beta}$ is a generalized Fisher matrix, which 
is related to the original Fisher matrix by
\begin{eqnarray}
\label{Falt}
F'_{\alpha\beta} = F_{\alpha\beta} + M^{-1}_{\alpha\beta}
\end{eqnarray}
and the maximum likelihood estimate $\hat p_\alpha^{\rm max} = \bar p_\alpha + \delta \hat p_\alpha^{\rm max}$
is given by (superseding eq. [\ref{dpmax}]):
\begin{eqnarray}
\label{dpmax2}
\delta \hat p_\alpha^{\rm max} \equiv 
\sum_{\beta} 
{F'}^{-1}_{\alpha\beta} 
\Bigl[ \sum_{i,j} {\partial d_i \over \partial p_\beta} C^{-1}_{ij} (\hat d_j - \bar d_j) \\ \nonumber
+ \sum_\gamma M^{-1}_{\beta\gamma} (\hat p_\gamma - \bar p_\gamma) \Bigr]
\end{eqnarray}

Therefore, the expectation value of $\hat p_\alpha^{\rm max}$ is given by
\begin{eqnarray}
\label{pave2}
\langle \hat p_\alpha^{\rm max} \rangle = \bar p_\alpha + \sum_{\beta} 
{F'}^{-1}_{\alpha\beta} 
\Bigl[ \sum_{i,j} {\partial d_i \over \partial p_\beta} C^{-1}_{ij} (\langle \hat d_j \rangle - \bar d_j) \\ \nonumber
+ \sum_\gamma M^{-1}_{\beta\gamma} (\langle \hat p_\gamma \rangle - \bar p_\gamma) \Bigr]
\end{eqnarray}

A set of priors that are unbiased implies $\langle \hat p_\gamma \rangle = p_\gamma^{\rm true}$.
The estimation bias is given by $\langle \hat p_\alpha^{\rm max} \rangle - p_\alpha^{\rm true}$,
and the error covariance is given by the inverse of the generalized Fisher matrix:
\begin{eqnarray}
\label{pperror2}
\langle \hat p_\alpha^{\rm max} \hat p_\beta^{\rm max}\rangle - \langle \hat p_\alpha^{\rm max} \rangle
\langle \hat p_\beta^{\rm max} \rangle 
= {F'}^{-1}_{\alpha\beta}
\end{eqnarray}

Eq. [\ref{pave2}] and [\ref{pperror2}] constitute the main results of this Appendix.
Let us illustrate their use by applying them to a question of interest in this paper:
what is the resulting bias on cosmological parameters if one ignores magnification bias
when interpreting ISW measurements? The parameters $p_\alpha$ stand for $w$, $\Omega_m$ and
so on. The data $\langle \hat d_j \rangle$ are the $C^{gT} + C^{\mu T}$ at various
$\ell$'s and redshifts (collectively labeled by the index $j$ here).
The $C_{ij}$ matrix here is the covariance described in eq. [\ref{covmat}].
To figure out the estimation bias on parameter $p_\alpha$ (which in our case is
mainly the dark energy equation of state $w$ since all other parameters
are fairly constrained by our priors), one applies eq. [\ref{pave2}] 
with $F'_{\alpha\beta}$, $\partial d_i/\partial p_\beta$, $C_{ij}$ and $\bar d_j$
all computed ignoring magnification bias. To be completely explicit, eq. [\ref{pave2}]
tells us that the average inferred
value for parameter $p_\alpha$ (e.g. $p_1 = w$ and so on) is:
\begin{eqnarray}
\label{wbias}
p_\alpha^{\rm inferred} = && \bar p_\alpha + \sum_{\beta} {F'^g}^{-1}_{\alpha\beta} \\ \nonumber
&& \Bigl[ \sum_{\ell,i,j} {\partial C^{gT}_\ell(z_i) \over \partial p_\beta} [{\,\rm Cov}^g_\ell]^{-1}_{ij} 
(C^{nT}_\ell (z_j) - C^{gT}_\ell (z_j)) \\ \nonumber
&& + \sum_{\gamma} M^{-1}_{\beta\gamma} (p_\gamma^{\rm true} - \bar p_\gamma) \Bigr]
\end{eqnarray}
where $F'^g$ is the generalized Fisher matrix (eq. [\ref{Falt}]) (the superscript $g$ is
supposed to remind us that magnification is ignored) i.e.
\begin{eqnarray}
\label{Fprimeg}
{F'^g}_{\alpha\beta} = M^{-1}_{\alpha\beta}+\sum_{\ell,i,j} {\partial C_\ell^{gT} (z_i) \over \partial p_\alpha} [{\,\rm Cov\,}^{g}_\ell ]^{-1}_{ij} 
{\partial C_\ell^{gT} (z_j)\over \partial p_\beta}
\end{eqnarray}
and the covariance (with magnification ignored) is:
\begin{eqnarray}
\label{covg}
&& [{\,\rm Cov}^g_\ell]_{ij} = \\ \nonumber
&& \quad {(C_\ell^{gg}(z_i,z_j) + {\delta_{ij}/ n_i})C^{TT}_\ell + C_\ell^{gT}(z_i) C_\ell^{gT} (z_j)
\over f_{\rm sky} (2\ell + 1)}
\end{eqnarray}
where $n_i$ is the mean number of galaxies per unit area in redshift bin $i$.
Note that the covariance given above is not exact (nor is the implicit assumption that
the different $\ell$ modes are uncorrelated), but it should be fairly accurate for modes with $\ell \gsim 10$
(recall that we use $f_{\rm sky} = 0.5$ in our worked examples),
which dominate the signal-to-noise. The same statement can be made about the Fisher matrix.

The fiducial parameter values $\bar p_\alpha$'s 
should be chosen to be close to the {\it perceived} best--fit values
--- recall that 
to a modeler who ignores magnification, the {\it perceived} best--fit differs
from the true values.
All quantities on the right hand side of eq. [\ref{wbias}],
such as $F'^g$, $C^{gT}_\ell$ and the covariance, should be evaluated with
the parameters set at the fiducial values $\bar p_\alpha$'s.
The only exception is $C^{nT}_\ell (z_j)$ which represents the average {\it observed}
cross-correlation and should of course be computed using the true parameter values
(e.g. $w = -1$ and so on in our thought experiment in \S \ref{bias}).
In principle, the {\it perceived} best-fit values can differ quite a bit from the true values.
We therefore apply eq. [\ref{wbias}] iteratively: starting
with $\bar p_\alpha = p_\alpha^{\rm true}$, we use eq. [\ref{wbias}] to compute
$p_\alpha^{\rm inferred}$, and then setting $\bar p_\alpha = p_\alpha^{\rm inferred}$,
we reapply eq. [\ref{wbias}] and so on. Generally, convergence is achieved with
only a few iterations.

The estimation-bias of interest is $p_\alpha^{\rm inferred} - p_\alpha^{\rm true}$.
The associated (theoretical/model-dependent) error covariance 
is given by the inverse of the
generalized Fisher matrix ${F'^g}$ (eq. [\ref{Fprimeg}]) that ignores
magnification, and is evaluated at the same
$\bar p_\alpha$'s as above.

If one is instead interested in making error forecasts when magnification bias 
is taken into
account (as in \S \ref{forecasts}), 
one should use eq. [\ref{pperror2}] with a $F'_{\alpha\beta}$ that is computed
without ignoring magnification bias i.e.
\begin{eqnarray}
\label{Fprimen}
{F'}_{\alpha\beta} = M^{-1}_{\alpha\beta}+\sum_{\ell,i,j} {\partial C_\ell^{nT} (z_i) \over \partial p_\alpha} [{\,\rm Cov\,}_\ell ]^{-1}_{ij} 
{\partial C_\ell^{nT} (z_j)\over \partial p_\beta}
\end{eqnarray}
where the covariance is given by eq. [\ref{covmat}].

\newcommand\spr[3]{{\it Physics Reports} {\bf #1}, #2 (#3)}
\newcommand\sapj[3]{ {\it Astrophys. J.} {\bf #1}, #2 (#3) }
\newcommand\sapjs[3]{ {\it Astrophys. J. Suppl.} {\bf #1}, #2 (#3) }
\newcommand\sprd[3]{ {\it Phys. Rev. D} {\bf #1}, #2 (#3) }
\newcommand\sprl[3]{ {\it Phys. Rev. Letters} {\bf #1}, #2 (#3) }
\newcommand\np[3]{ {\it Nucl.~Phys. B} {\bf #1}, #2 (#3) }
\newcommand\smnras[3]{{\it Monthly Notices of Royal
        Astronomical Society} {\bf #1}, #2 (#3)}
\newcommand\splb[3]{{\it Physics Letters} {\bf B#1}, #2 (#3)}
\newcommand\AaA{Astron. \& Astrophys.~}
\newcommand\apjs{Astrophys. J. Suppl.}
\newcommand\aj{Astron. J.}
\newcommand\mnras{Mon. Not. R. Astron. Soc.~}
\newcommand\apjl{Astrophys. J. Lett.~}
\newcommand\etal{{\it et al.}}


\begin{thebibliography}{99}
\bibitem{gands} Strictly speaking, the factor of $2.5s - 1$ should be absorbed into $g$ 
where $s$ is made a function of the integration variable $z'$.
Here, we make the approximation that $s$ is roughly constant within the redshift bin $i$.
\bibitem{caveat} There is a slight caveat to this statement. If $s<0.4$, $C^{\mu T}$ will be negative. For a given $z_i$, $C^{\mu T}_{\ell}(z_i)$ is peaked at lower $\ell$ than $C^{gT}_{\ell}(z_i)$, thus for $s<0.4$ the peak of the sum $C^{nT}_{\ell}(z_i)=C^{gT}_{\ell}(z_i)+C^{\mu T}_{\ell}(z_i)$ is at higher $\ell$ than it would be if there were no magnification. So, in principle if the magnification term, $C^{\mu T}$ were large and negative the range of largest signal to noise may be shifted beyond $\ell=100$. However this is unlikely in practice because magnification bias does not become important until high redshift ($z\gsim 2$), and at high redshifts one is typically observing bright objects which tend to have $s>0.4$, unless one has an exceptionally deep survey.
\bibitem{comments} Strictly speaking, for Figures \ref{dSN} and \ref{dwerr},
the lines should really be represented as points, since the quantities plotted are fundamentally
derived from discrete redshift bins. We join them up as lines for clarity of presentation.
For Figures \ref{SN} and \ref{werr}, there is one good justification for connecting
the dots: in the limit the redshift bins become smaller and smaller (and the number of bins
increases), the cumulative signal-to-noise or the cumulative error should approach a smooth
curve approximated by the one we present.
\bibitem{lines} Strictly speaking, for Figure \ref{wtgData}, the dotted lines should really
be represented as points, since $w_{nT}/b$ has a residual dependence on the bias $b$, and
the bias is fundamentally sample specific and does not necessarily evolve in a smooth manner. 
We join up the points as lines for clarity of presentation.
\end{thebibliography}
\end{document}